\documentclass[twoside,twocolumn,9pt]{article}
\usepackage{extsizes}
\usepackage[super,sort&compress,comma]{natbib} 
\usepackage[version=3]{mhchem}
\usepackage[left=1.5cm, right=1.5cm, top=1.785cm, bottom=2.0cm]{geometry}
\usepackage{balance}
\usepackage{times,mathptmx}
\usepackage{sectsty}
\usepackage{graphicx} 
\usepackage{lastpage}
\usepackage[format=plain,justification=justified,singlelinecheck=false,font={stretch=1.125,small,sf},labelfont=bf,labelsep=space]{caption}
\usepackage{float}
\usepackage{fancyhdr}
\usepackage{fnpos}
\usepackage[english]{babel}
\addto{\captionsenglish}{%
  
}
\usepackage{array}
\usepackage{droidsans}
\usepackage{charter}
\usepackage[T1]{fontenc}
\usepackage[usenames,dvipsnames]{xcolor}
\usepackage{setspace}
\usepackage[compact]{titlesec}
\usepackage{hyperref}
\usepackage{epstopdf}%This line makes .eps figures into .pdf - please comment out if not required.

\definecolor{cream}{RGB}{222,217,201}

\begin{document}

\pagestyle{fancy}
\thispagestyle{plain}
\fancypagestyle{plain}{

%%%HEADER%%%
\fancyhead[C]{\includegraphics[width=18.5cm]{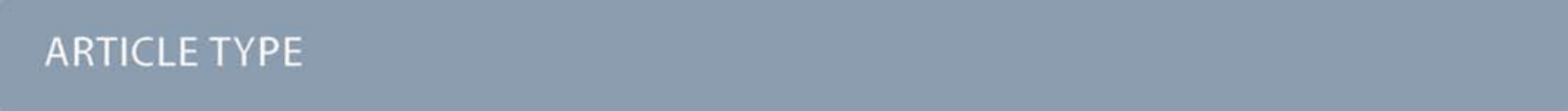}}
\fancyhead[L]{\hspace{0cm}\vspace{1.5cm}\includegraphics[height=30pt]{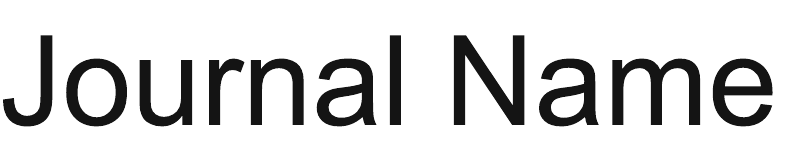}}
\fancyhead[R]{\hspace{0cm}\vspace{1.7cm}\includegraphics[height=55pt]{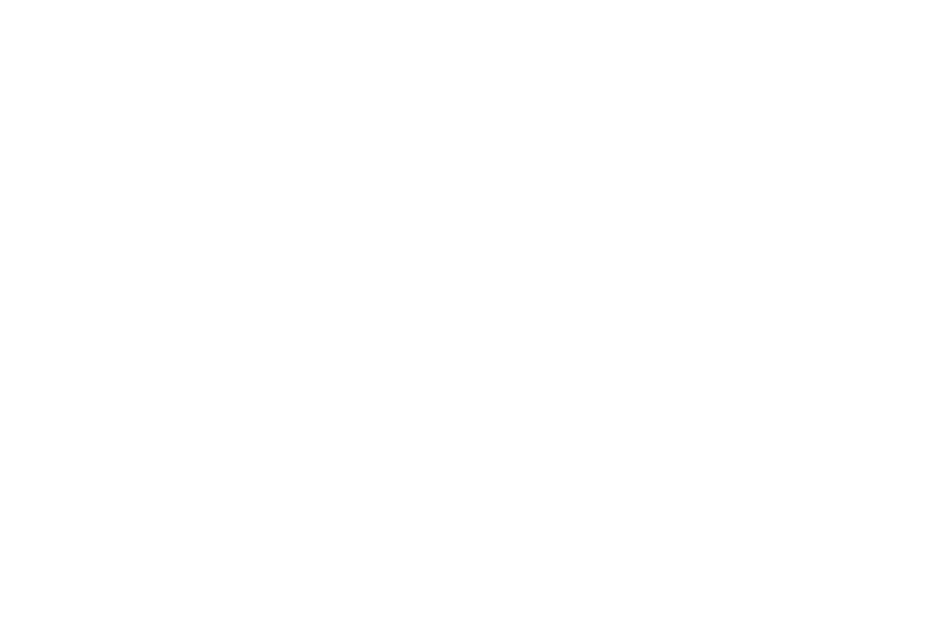}}
\renewcommand{\headrulewidth}{0pt}
}
%%%END OF HEADER%%%

%%%PAGE SETUP - Please do not change any commands within this section%%%
\makeFNbottom
\makeatletter
\renewcommand\LARGE{\@setfontsize\LARGE{15pt}{17}}
\renewcommand\Large{\@setfontsize\Large{12pt}{14}}
\renewcommand\large{\@setfontsize\large{10pt}{12}}
\renewcommand\footnotesize{\@setfontsize\footnotesize{7pt}{10}}
\makeatother

\renewcommand{\thefootnote}{\fnsymbol{footnote}}
\renewcommand\footnoterule{\vspace*{1pt}% 
\color{cream}\hrule width 3.5in height 0.4pt \color{black}\vspace*{5pt}} 
\setcounter{secnumdepth}{5}

\makeatletter 
\renewcommand\@biblabel[1]{#1}            
\renewcommand\@makefntext[1]% 
{\noindent\makebox[0pt][r]{\@thefnmark\,}#1}
\makeatother 
\renewcommand{\figurename}{\small{Fig.}~}
\sectionfont{\sffamily\Large}
\subsectionfont{\normalsize}
\subsubsectionfont{\bf}
\setstretch{1.125} %In particular, please do not alter this line.
\setlength{\skip\footins}{0.8cm}
\setlength{\footnotesep}{0.25cm}
\setlength{\jot}{10pt}
\titlespacing*{\section}{0pt}{4pt}{4pt}
\titlespacing*{\subsection}{0pt}{15pt}{1pt}
%%%END OF PAGE SETUP%%%

%%%FOOTER%%%
\fancyfoot{}
\fancyfoot[LO,RE]{\vspace{-7.1pt}\includegraphics[height=9pt]{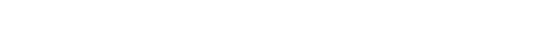}}
\fancyfoot[CO]{\vspace{-7.1pt}\hspace{13.2cm}\includegraphics{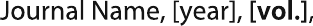}}
\fancyfoot[CE]{\vspace{-7.2pt}\hspace{-14.2cm}\includegraphics{head_foot/RF}}
\fancyfoot[RO]{\footnotesize{\sffamily{1--\pageref{LastPage} ~\textbar  \hspace{2pt}\thepage}}}
\fancyfoot[LE]{\footnotesize{\sffamily{\thepage~\textbar\hspace{3.45cm} 1--\pageref{LastPage}}}}
\fancyhead{}
\renewcommand{\headrulewidth}{0pt} 
\renewcommand{\footrulewidth}{0pt}
\setlength{\arrayrulewidth}{1pt}
\setlength{\columnsep}{6.5mm}
\setlength\bibsep{1pt}
%%%END OF FOOTER%%%

%%%FIGURE SETUP - please do not change any commands within this section%%%
\makeatletter 
\newlength{\figrulesep} 
\setlength{\figrulesep}{0.5\textfloatsep} 

\newcommand{\topfigrule}{\vspace*{-1pt}% 
\noindent{\color{cream}\rule[-\figrulesep]{\columnwidth}{1.5pt}} }

\newcommand{\botfigrule}{\vspace*{-2pt}% 
\noindent{\color{cream}\rule[\figrulesep]{\columnwidth}{1.5pt}} }

\newcommand{\dblfigrule}{\vspace*{-1pt}% 
\noindent{\color{cream}\rule[-\figrulesep]{\textwidth}{1.5pt}} }

\makeatother
%%%END OF FIGURE SETUP%%%

%%%TITLE, AUTHORS AND ABSTRACT%%%
\twocolumn[
  \begin{@twocolumnfalse}
\vspace{3cm}
\sffamily
\begin{tabular}{m{4.5cm} p{13.5cm} }

\includegraphics{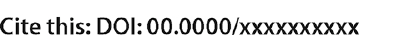} & \noindent\LARGE{\textbf{Role of precursor composition in the polymorph transformations, morphology control and ferromagnetic properties of nanosized TiO\textsubscript{2}}} \\%Article title goes here instead of the text "This is the title"
\vspace{0.3cm} & \vspace{0.3cm} \\

 & \noindent\large{Dmitry Zablotsky,$^{\ast}$\textit{$^{a}$} Mikhail M. Maiorov,\textit{$^{a}$} Aija Krumina,\textit{$^{a,b}$} Marina Romanova,\textit{$^{c}$} and Elmars Blums\textit{$^{a}$}} \\%Author names go here instead of "Full name", etc.

\includegraphics{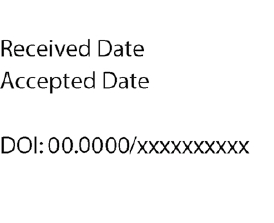} & \noindent\normalsize{Pure phase and mixed phase TiO$_2$ nanoparticles have been produced using a pyrolytic method from a non-aqueous carboxylate precursor. The precursor was prepared by a multiphase cation exchange using pentanoic acid (C$_4$H$_9$COOH). The thermal stability, polymorph content, morphology, size distribution and surface region of the produced nanoparticles were studied by TGA/DSC, XRD, FTIR and TEM. High quality monodisperse nanoparticles have been produced in the size range from 7 to 27 nm. The nanoparticles showed room temperature ferromagnetism revealed by VSM within bound polaron model. The carboxylate precursor is a good alternative to standard sol-gel to produce nanoparticles free from impurities.} \\%The abstrast goes here instead of the text "The abstract should be..."

\end{tabular}

 \end{@twocolumnfalse} \vspace{0.6cm}

  ]
%%%END OF TITLE, AUTHORS AND ABSTRACT%%%

%%%FONT SETUP - please do not change any commands within this section
\renewcommand*\rmdefault{bch}\normalfont\upshape
\rmfamily
\section*{}
\vspace{-1cm}

%%%FOOTNOTES%%%

\footnotetext{\textit{$^{a}$~University of Latvia, Jelgavas 3, 1004 Riga, Latvia; E-mail: dmitrijs.zablockis@lu.lv}}

\footnotetext{\textit{$^{b}$~Institute of Inorganic Chemistry, Riga Technical University, Faculty of Material Science and Applied Chemistry, P. Valdena 7, 1048 Riga, Latvia.}}

\footnotetext{\textit{$^{c}$~Institute of Biomedical Engineering and Nanotechnologies, Riga Technical University, Viskalu 36A, 1006 Riga, Latvia.}}

%%%END OF FOOTNOTES%%%

%%%MAIN TEXT%%%%

\section{Introduction}

Titanium dioxide (TiO$_2$) is a multifunctional semiconducting metal oxide (SMO) that is expected to play a significant role in environmental and energy applications \cite{Shayegan2018,Byrne2018}.
Amongst all SMOs (such as WO$_3$, ZnO, Fe$_2$O$_3$, MnO$_2$ etc.) TiO$_2$ is the most intensively studied because its chemically inert, reusable, abundant and inexpensive, non-toxic and biologically compatible and has a wide band gap ($\sim$3 eV), which makes it an ideal photocatalyst that can be used to break down and mineralize organic pollutants, inhibit bacterial growth, self-sterilize surfaces or split water to hydrogen and oxygen by solar light\cite{Ali2018}.
In the past decades a tremendous effort has been put into improving the efficiency of TiO$_2$ materials. Titania crystallizes in three distinct polymorphs with different properties and structure: rutile (tetragonal, space group P42/mmm), anatase (tetragonal, space group I41/amd) and brookite (orthorhombic, space group Pcab). Bi-phasic anatase/rutile mixtures are most wanted than pure phases because of superior properties \cite{Wu2004, Kho2010, Li2009, Li2007, Bacsa1998}. The established standard - Aeroxide TiO$_2$ P25 marketed by Evonik Industries (Degussa) – a mixed phase photocatalyst consisting of about 70-80 wt.\% anatase and 20-30 wt.\% rutile, has been extensively studied, benefited from synergistic interaction of the phases \cite{Jiang2018, Bickley1991, Kawahara2002, Hurum2005, Ohtani2010}.
Fundamentally, a considerable problem is a very limited range of precursors and process chemistries to produce nanostructured TiO$_2$. The solution-based sol–gel chemistry is commonly used to produce TiO$_2$ nanocrystallites. It starts from either an organometallic titanium alkoxide (Ti ethoxide, butoxide or isopropoxide) or inorganic titanium chloride (TiCl$_4$ or TiCl$_3$) and proceeds via a two-step process: hydrolysis of the salt by added water resulting in the formation of intermediate species (monomers) and their subsequent self-assembly and polymerization into extended 3D network. The colloidal suspension (sol) is precipitated, dried and calcined to complete the phase transformation.

Previous studies have shown that during sol-gel the phase content, crystalline structure, size and morphology of particles were found to be highly sensitive to various parameters, such as the Ti:H$_2$O ratio, anion content (e.g. Cl$^-$), the pH, and the choice of reaction modulator \cite{Goutailler2002, Yu2003, Bessekhouad2003, Isley2006, Ranade2002, Liu2008, DiPaola2008}. The hydrolysis of the Ti precursor is a critical step, which determines the initial monomer species and strongly impacts the resultant phase \cite{Isley2006, Cihlar2015}. For instance, a change of molar ratios in the chemical environment (such as decrease in instantaneous concentration of Ti precursor) results in different hydrolysis speeds and time-dependent shifts of polymorph selectivity \cite{Smonarson2019}. The complexity of the method and the lack of detailed understanding of the chemical equilibrium of the species in solution and kinetics of nucleation and growth of the different phases makes it difficult to achieve and reproduce a substantial control over the relative phase content, usually obtaining a final mixture of anatase, rutile and brookite.
Secondly, there are safety concerns over direct use of Ti alkoxides or chlorides in sol-gel as they are in general very reactive and sensitive to moisture. The hydrolysis is vigorous (even at 0 $^\circ$C), hence, extensive precautions are necessary in handling these chemicals. Likewise, the sol-gel generates highly corrosive acidic conditions, since the alkoxide route requires acid catalysts (e.g. HCl, HNO$_3$, H$_2$SO$_4$ \cite{Isley2006, DiPaola2008, Smonarson2019}) to control the steps in the reaction and phase selectivity, whereas chloride precursors (TiCl$_4$ or TiCl$_3$) produce large amounts of corrosive HCl \textit{in situ} during hydrolysis, which results in chloride-rich solutions with very low
pH.
Moreover, extensive dialysis/purification procedures are required to remove the byproducts of sol-gel synthesis adsorbed on the particle surface and strongly bound (e.g. Cl$^-$, NO$_3^-$, NH$_4^+$) species obtained from precursors cannot be totally removed \cite{Liu2008, Pottier2003}. The doping impurities, secondary impurity phases and surface-adsorbed features definitely influence the surface properties, which can hinder the application potential of the material.
For instancce, Aeroxide (Degussa) P25 - the most successful and used photocatalyst - is instead produced by pyrogenic flame-hydrolysis, in which vapourised TiCl$_4$ is combusted in an oxy-hydrogen flame \cite{degussa_patent}. The solid is then separated and treated with steam at 450-550 $^\circ$C to remove chlorine-containing groups. The main advantage of pyrohydrolysis is that it is scalable to industrial levels. Nevertheless, a toxic chlorine-rich flame presents significant corrosive hazards and is unfriendly to the environment. Beyond the obvious concerns with environmental pollution, Ti chlorides are not easy to store and handle. 
Thus, it is a practically and scientifically important challenge to develop alternative precursors to control the size, polymorph content and morphology of TiO$_2$ nanocrystals and, therefore, optimize the properties of material. 
A metal-organic precursor is more favourable than inorganic titanium chlorides, because it can be used in a chlorine-free system without related hazards, however, the widely used Ti alkoxides suffer from instability and poor cost-efficiency.

In this study we report the development of substitute precursors to produce nanostructured TiO$_2$ via pyrolysis or thermal decomposition using metal-organic carboxylate-based extraction systems. Having the advantage of being inexpensive and environmentally friendly, metal carboxylate complexes have been used extensively in the production of nanoparticles of SMOs \cite{Yu2002,Yu2003b,Yu2004,Park2004}, but the production of TiO$_2$ nanoparticles from a carboxylate precursor has not been reported. Herein, Ti chloride is used as the initial titanium source preserving the key advantage of low cost, however, the chloride anions Cl$^-$ are avoided with the extraction system via aqueous/organic phase segregation and metal cation exchange resulting in \textit{in situ} dialysis. The resulting precursors are storable and the nanostructured TiO$_2$ is produced by non-hydrolytic pyrolysis. Thus, we are able to overcome many of the specific problems described above to achieve better particle size and polymorph control than hydrolytic routes by simply varying the temperature of pyrolysis.

\section{Experimental}

\subsection{Preparation of Ti-containing organometallic precursor and pyrolization treatment}

Initially, an aqueous solution of TiCl$_3$ was produced by dissolving fine-grained c.p. titanium powder (1.2 g, particle size 63-100 $\mu$m, -140+230 mesh) in boiling hydrochloric acid solution (60 ml, 1:1 vol.). The metal concentration was adjusted to 0.1M by dH$_2$O addition (pH 0.5). Titanium-rich carboxylate-based organometallic precursor was produced by liquid-liquid extraction: 60 ml of TiCl$_3$ aqueous solution and 20 ml (3:1 aqueous/organic volume ratio) of pentanoic acid (without diluent) were placed in a separation funnel; 1M NaOH soluion was added to the mixture in small portions, followed by vigorous shaking after each addition until the pH of the aqueous phase reached pH 1, after which the phases were allowed to separate for 0.5 h. After complete stratification and removal of the aqueous phase followed by filtration, a Ti-rich precursor E1 (0.15M Ti) was produced.
To study the effect of precursor concentration on produced TiO$_2$ a more concentrated precursor E2 (0.5M Ti) was prepared as described above, but with 5:1 aqueous/organic volume ratio. The precursors E1 and E2 were pyrolyzed for 1 h at 350 $^\circ$C (E1/E2-350), 400 $^\circ$C (E1/E2-400), 450 $^\circ$C (E1/E2-450), 550 $^\circ$C (E1/E2-550), 650 $^\circ$C (E1/E2-650), 750 $^\circ$C (E1/E2-750) in static air to sample the phase content of derived TiO$_2$ nanoparticles \cite{Serga2018}.

\begin{figure}
	\centering
		\includegraphics[width=1\linewidth]{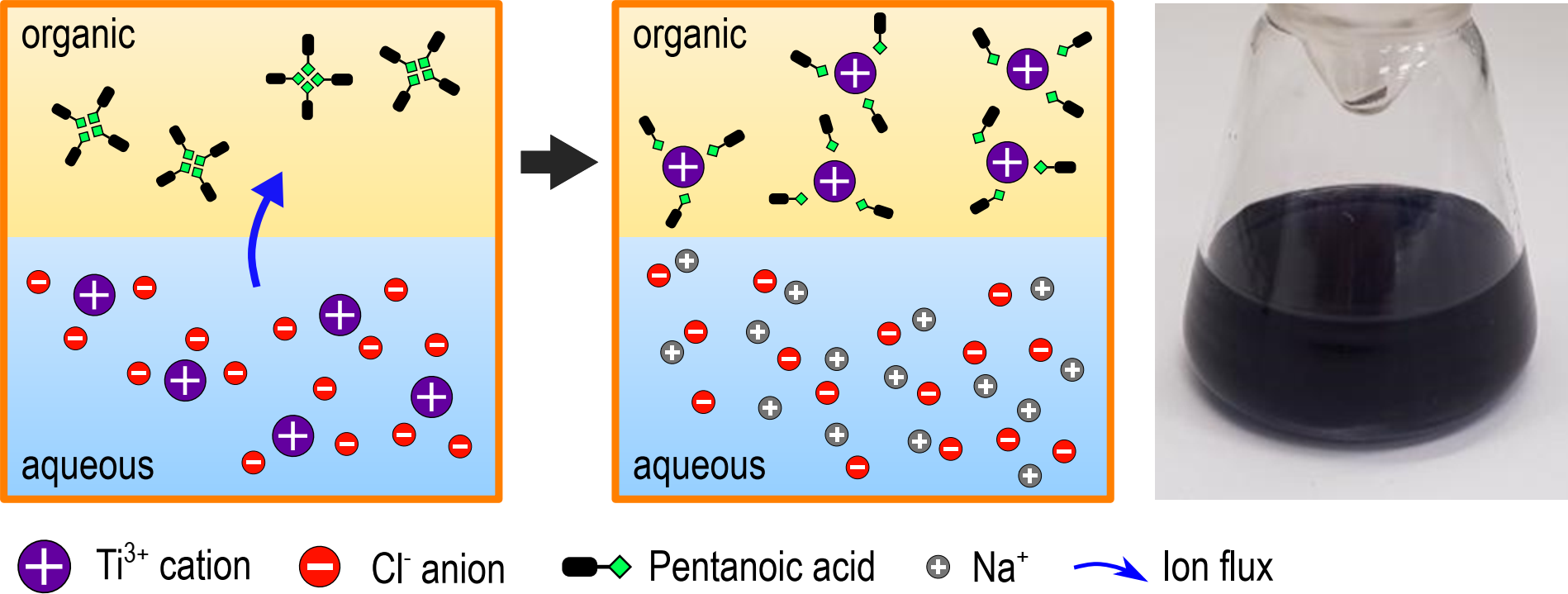}
	\caption{ Schematic of preparation of carboxylate-based Ti organometallic precursor: Left - phase contact of TiCl$_3$ aqueous solution and pentanoic acid, middle - extraction of Ti$^{3+}$ from the aqueous solution into the organic phase by cation exchange, right - produced non-aqueous Ti$^{3+}$-rich carboxylate-based organometallic precursor with characteristic violet color. }
	\label{FIG:1}
\end{figure}

\subsection{Preparation of reference TiO$_2$ by sol-gel}

A reference sample was prepared by standard sol-gel method, using 250 ml of 0.1M TiCl$_3$ aqueous solution produced in the previous step. The gel-like precipitate was obtained by dropwise addition of 0.5M NaOH aqueous solution at a rate of 3 ml/min under vigorous stirring until pH 6 of the solution was reached. The produced gel was aged for 24 h at ambient conditions, followed by filtration and washing with dH$_2$O. The absence of chlorine in the decantate was checked by AgNO$_3$ solution. The as-dried precipitate (P) was calcined at 450 $^\circ$C (P-450), 550 $^\circ$C (P-550), 650 $^\circ$C (P-650), 750 $^\circ$C (P-750) in static air to produce TiO$_2$ nanoparticles.

\subsection{Characterization}

The thermal stability of the produced precursors was studied by thermal gravimetric analysis (TGA) and differential scanning calorimetry (DSC) using the STA PT1600 (LINSEIS) device. The sample was heated in static air from ambient temperature to 700$^\circ$C at a rate of 10 $^\circ$C/min.

IR spectra were recorded at r.t. using Bruker Tensor II FT-IR spectrophotometer. For each spectrum 36 scans were performed in the range 4000-400 cm$^{-1}$ with 7 mm KBr discs (TiO$_2$/KBr mass ratio 1:100) prepared under a load of 2000 kg.

Powder X-ray diffraction (XRD) patterns of dried samples were recorded in $2\theta$ range between 20$^\circ$ and 75$^\circ$ using D8 Advance (Bruker Corporation) diffractometer with CuK$\alpha$ radiation ($\lambda$ = 1.5418\AA), accelerating voltage 40 kV and current 40 mA; step size 0.02$^\circ$ at a scanning rate 2 s/step. Zero-background Si sample holders were used. The quantitative phase composition of the produced materials were determined by Rietvield refinement (FullProf Suite ver. 7.30), a whole pattern fitting method that systematically varies constraints in a simulated pattern to reproduce the experimental pattern. The refinement was performed starting from the identified crystal phases and known crystal structure data using isotropic size broadening (as confirmed by HRTEM snapshots), Pseudo-Voigt reflection profile approximation with Caglioti equation (FWHM$^2=U\tan^2\theta + V\tan\theta+W$) and experimental instrumental broadening parameters with standard reference material. The background was modeled using a 6th-order shifted polynomial. The parameters refined were zero shift, scale factors, unit cell and peak shape.

The morphology of the produced nanoparticles was checked by transmission electron microscopy (TEM) (FEI Technai G2 F20 operating at 200 kV) in bright field mode. Dynamic light scattering (DLS, Malvern Instruments, Zetasizer Nano S90) was used to measure the hydrodynamic size distribution in particle suspensions produced by electric ablation. For the measurement the initial suspension was diluted with distilled water to achieve optimum count rate.

The ferromagnetic properties of the produced samples were investigated by vibrating sample magnetometry (VSM). The magnetization curve was recorded at r.t. (Lake Shore Cryotronics instrument 7404VSM) in a field range up to 1 T. Several full magnetization cycles have been performed  to detect potential hysteretic behavior.

\section{Results and discussion}

The preparation of reference sample (P) within a standard sol-gel procedure involves the hydrolysis of TiCl$_3$, which generates highly acidic (low pH) and chloride (Cl$^-$) rich solution. The selectivity of formed TiO$_2$ polymorphs is highly dependent on instantaneous pH, Ti concentration and reaction time \cite{Pottier2003,DiPaola2008}, where small change in molar ratios would result in different phase structure \cite{Wang2014}. At high Ti concentration ([Ti] > 0.25 M) the solution will predominantly contain [TiO(OH$_2$)$_5$]$^{2+}$ monomers and the condensation will proceed by olation by sharing equatorial edges \cite{DiPaola2009} with formation of hydroxo bridges (Ti-OH-Ti) \cite{Smonarson2019}. In this case, rutile-rich crystallites are precipitated \cite{DiPaola2009, Roca2012, ReyesCoronado2008, Abbasi2015, Zhou2016}. To stabilize the production of anatase, Ti concentration was decreased to 0.1 M \cite{Pottier2003, DiPaola2009, Smonarson2019}. Moreover, the pH was raised to 6 to avoid the formation of a substantial amount of brookite \cite{Pottier2003, DiPaola2009, Zhou2016}. Conversion of the amorphous hydrated precipitate (P) to crystalline anatase is realized by calcination treatment.

The organometallic Ti precursors E1 and more concentrated E2 were produced by liquid-liquid extraction: the extraction of Ti$^{3+}$ from aqueous phase (aq) into the organic
phase (o) by pentanoic acid is realized by cation exchange (Fig.~\ref{FIG:1}), according to equation
\begin{equation}
Ti^{3+}\left(aq\right)+3HOOCR\left(o\right) \leftrightarrow Ti\left(OOCR\right)_3\left(o\right)+3H^+\left(aq\right) \nonumber
\end{equation}
Pentanoic acid is an efficient cation exchanger due to its favourable hydrophylic/hydrophobic balance. The organometallic extracts, which were brightly violet immediately after extraction (Fig.~\ref{FIG:1}), decolorated shortly after production due to the change of the Ti oxidation state from Ti$^{3+}$ to Ti$^{4+}$ and were long-term storable.

\subsection{Thermal analysis}

The thermal behaviour of the precursors, phase transition and phase stability of produced TiO$_2$ nanoparticles were studied using TGA-DSC analysis. The measured thermoanalytical curves in the range between 20 and 700 $^\circ$C are shown in Fig. 2. The thermogravimetric analysis of the as-dried reference precipitate (P) shows active weight loss (ca. 20wt.\%) when the sample is heated from r.t. to 200 $^\circ$C, which is accompanied by a pronounced endothermic peak (peak max at 90 $^\circ$C) in the DSC trace. This step is attributed to the loss of weakly bound physisorbed water from the surface of the particles. The second step in the range between 200 $^\circ$C and 400 $^\circ$C (peak max at 290 $^\circ$C) is identified by a broad endothermic process related to the liberation of crystallization water and chemisorbed surface hydroxyls from hydrated precipitates (titanium oxyhydrate). This is accompanied by a small mass loss ca. 5 wt.\%. According to the XRD results (see below) the crystallization of anatase from the precipitates starts at about 450 $^\circ$C, which does not associate to any DSC event. Further temperature increase results in prolonged weight loss of the precipitates at a reduced rate, associated with continuing dehydration process up to 1000 $^\circ$C.
\begin{figure}
	\centering
		\includegraphics[width=0.8\linewidth]{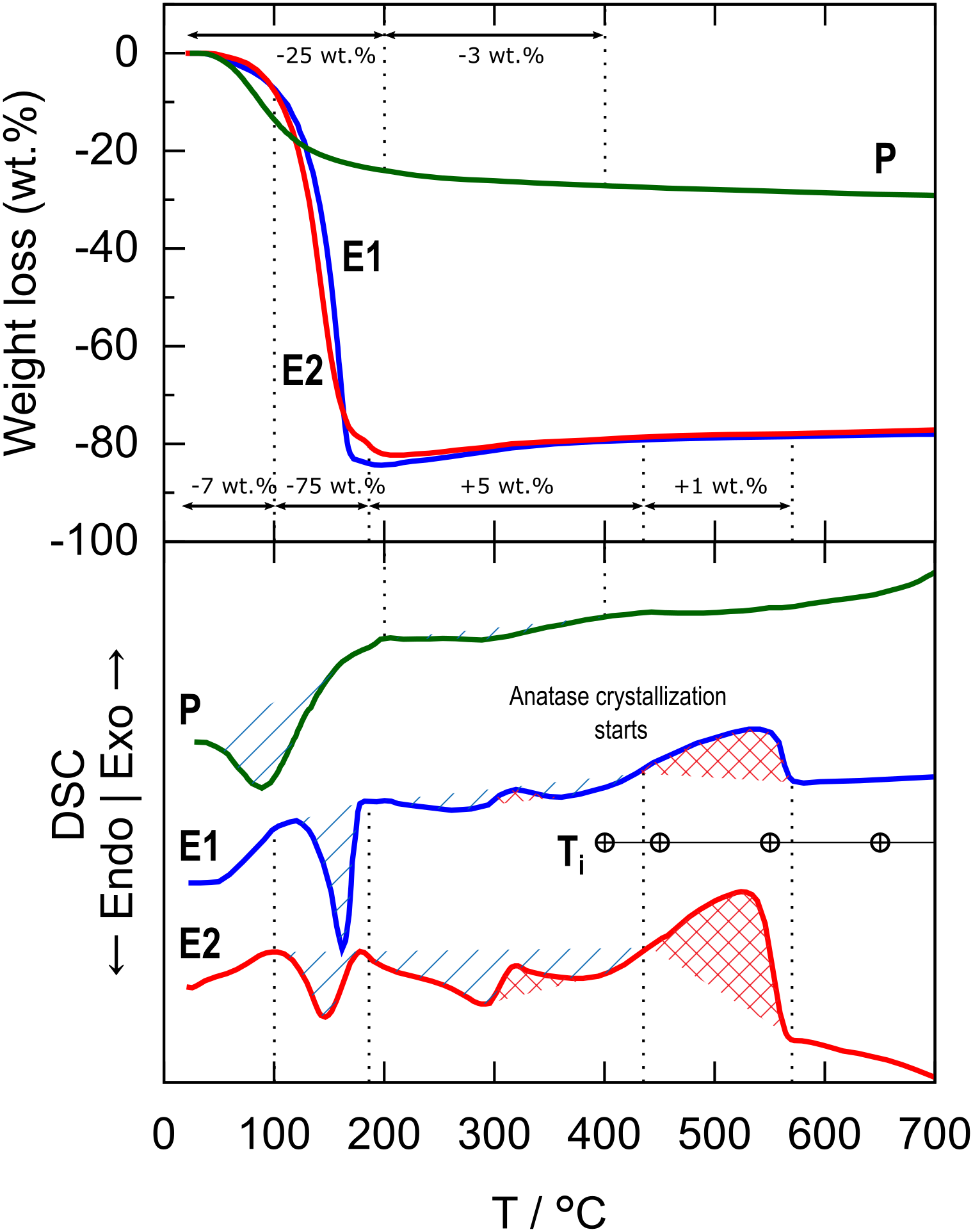}
	\caption{ TGA-DSC traces of carboxylate organometallic precursors E1 (0.15M Ti) and E2 (0.5M Ti) produced by liquid-liquid extraction and sol-gel precipitates P from monophasic precipitation. Symbols indicate sampling temperatures T$_i$, $^\circ$C: 350, 400, 450, 550, 650, 750 }
	\label{FIG:2}
\end{figure}
\begin{figure*}
	\centering
		\includegraphics[width=1\linewidth]{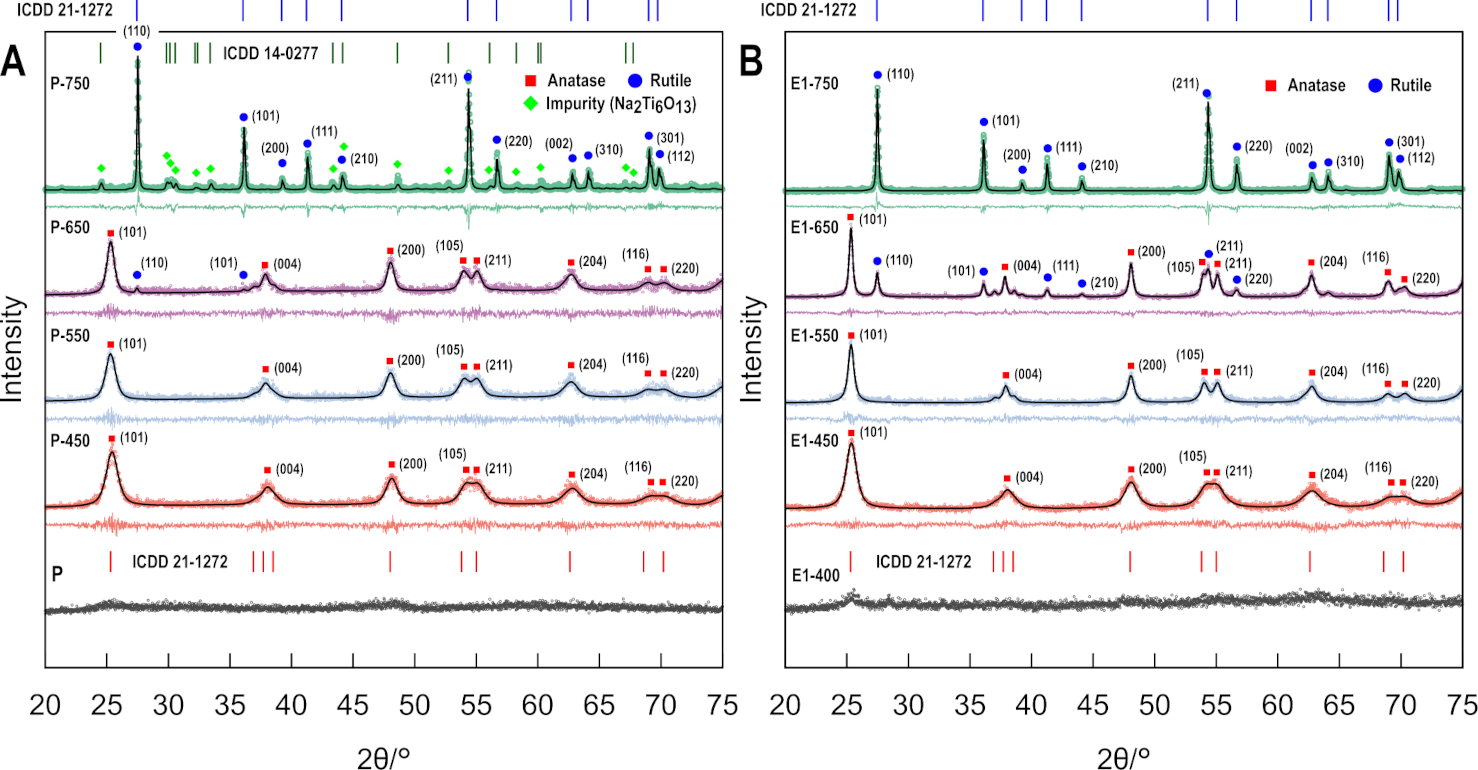}
	\caption{ X-ray diffraction patterns (symbols) and corresponding Rietvield-refined profiles (black lines): A - calcined precipitates produced by standard sol-gel method. B - TiO$_2$ samples produced by pyrolyzation of organometallic precursor. The residual lines show the difference of the measured and simulated XRD profiles. }
	\label{FIG:3}
\end{figure*}

The organometallic carboxylate Ti precursors E1 and E2 produced by liquid-liquid extraction show a strong endothermic peak in the range between 100 $^\circ$C and 186 $^\circ$C related to the active evaporation of free extractant (pentanoic acid) and is accompanied by a pronounced mass loss ca. 75 wt.\%, which terminates around the boiling point of pentanoic acid (T$_{boil}$=186 $^\circ$C). At this point the mass loss ends and amounts to 80-85 wt.\%. 
Further temperature increase initiates the thermal decomposition of non-reactive titanium pentanoate contained in the precursor, which should proceed according to the standard route for transition metal carboxylates \cite{ Rai1979, Seddon1986}: Ti(OOCR)$_4\rightarrow $TiO$_2$+2R$_2$CO+2CO$_2$. The decarboxylation of the organic component is accompanied by the formation of ketones containing 3 to 9 carbon atoms (i.e. dibutyl ketone, dipropyl ketone, diethyl ketone, acetone) \cite{Judd1974, Leicester2007, Seddon1986}. The ketones evolved during this process also undergo decomposition with atmospheric oxygen to yield CO$_2$ and H$_2$O \cite{Rai1979}. 
The third step in the temperature range between 200 $^\circ$C and ca. 450 $^\circ$C is marked by the broad endothermal process associated with breakdown of organic groups and release of volatile species, mainly CO$_2$ and formed ketones \cite{ Judd1974, Leicester2007, Seddon1986}. The signature of this process is more pronounced in the DSC trace of precursor E2 having higher Ti concentration. The liberation of the titanium ion from the organometallic precursor leads to the formation of amorphous nuclei of TiO$_2$ nanoparticles. This process is accompanied by the gradual increase of the sample mass (by 5-6 wt.\%) owing to the oxygen binding from the air during titanium oxidation. The exothermic event at ca. 320 $^\circ$C in the DSC trace indicates the energy release from combustion/burning of carbonaceous residues.
The more gradual removal of organic groups continues up to 450 $^\circ$C, as shown by infrared spectroscopy (see below), which indicates that the carboxylate (COO$^-$) functional group from pentanoic acid is fully desorbed at this temperature. TGA analysis shows that the conversion of titanium pentanoate to TiO$_2$ is complete at 450 $^\circ$C with no further mass change. As confirmed by XRD analysis (see below) the produced TiO$_2$ is amorphous giving no XRD trace of crystalline structure. The formation of high crystalline anatase starts at ca. 450 $^\circ$C accompanied by a broad exothermic process (peak max ca. 530 $^\circ$C) associated with restructuring and grain growth of anatase particles. The transformation is completed at ca. 560 $^\circ$C the product fully transforms into anatase phase. 

\begin{table*}
\small
\caption{Summary of sample phase structure refinement. A - anatase, R - rutile, Imp. - impurity phase Na$_2$Ti$_6$O$_{13}$}
\begin{tabular*}{\textwidth}{@{\extracolsep{\fill}}lrcrcrrcrlr}
\hline
T, $^\circ$C          & \multicolumn{5}{c}{Primary phase}                                                                                                               & \multicolumn{5}{c}{Secondary phase}                                                                                                                                         \\
              & \multicolumn{2}{c}{Unit cell, \AA} & \multicolumn{1}{l}{wt.\%} & \multicolumn{1}{l}{Phase} & \multicolumn{1}{c}{Crystallite}     & \multicolumn{2}{c}{Unit cell, \AA}                                                  & \multicolumn{1}{l}{wt.\%} & Phase                    & \multicolumn{1}{c}{Crystallite}     \\ \cline{2-3} \cline{7-8}
              & \multicolumn{1}{c}{a, b}  & \multicolumn{1}{c}{c} & \multicolumn{1}{l}{}      & \multicolumn{1}{l}{}      & \multicolumn{1}{c}{size, d$_{XRD}$} & \multicolumn{1}{c}{a, b}                               & \multicolumn{1}{c}{c} & \multicolumn{1}{l}{}      &                          & \multicolumn{1}{c}{size, d$_{XRD}$} \\
              \hline
\multicolumn{11}{l}{Sol-gel reference sample (P)}                                                                                                                                                                                                                                                                                            \\
r.t.          & \multicolumn{5}{c}{- amorphous -}                                                                                                               & \multicolumn{1}{l}{}                                   & \multicolumn{1}{l}{}  & \multicolumn{1}{l}{}      &                          & \multicolumn{1}{l}{}                \\
450           & 3.795                     & 9.492                 & 100\%                     & A                         & 7 nm                                &                                                        & \multicolumn{1}{r}{}  &                           &                          & \multicolumn{1}{l}{}                \\
550           & 3.793                     & 9.497                 & 100\%                     & A                         & 9 nm                                &                                                        & \multicolumn{1}{r}{}  &                           &                          & \multicolumn{1}{l}{}                \\
650           & 3.790                     & 9.504                 & 95\%                      & A                         & 12 nm                               & 4.593                                                  & 2.958                 & 5\%                       & \multicolumn{1}{c}{R}    & 18 nm                               \\
750           & 4.596                     & 2.960                 & 79\%                      & R                         & 64 nm                               & \begin{tabular}[c]{@{}r@{}}15.109\\ 3.749\end{tabular} & 9.181                 & 21\%                      & \multicolumn{1}{c}{Imp.} & 45 nm                               \\
\multicolumn{11}{l}{Pyrolyzed carboxylate precursor E1 (0.15M)}                                                                                                                                                                                                                                                                                       \\
350           & \multicolumn{5}{c}{- amorphous -}                                                                                                               & \multicolumn{1}{l}{}                                   &                       & \multicolumn{1}{l}{}      &                          & \multicolumn{1}{l}{}                \\
400           & \multicolumn{5}{c}{- amorphous -}                                                                                                               & \multicolumn{1}{c}{}                                   &                       & \multicolumn{1}{c}{}      & \multicolumn{1}{c}{}     &                                     \\
450           & 3.794                     & 9.481                 & 100\%                     & A                         & 8 nm                                &                                                        &                       &                           & \multicolumn{1}{c}{}     &                                     \\
550           & 3.790                     & 9.504                 & 100\%                     & A                         & 15 nm                               &                                                        &                       &                           & \multicolumn{1}{c}{}     &                                     \\
650           & 3.788                     & 9.513                 & 65\%                      & A                         & 24 nm                               & 4.596                                                  & 2.961                 & 35\%                      & \multicolumn{1}{c}{R}    & 27 nm                               \\
750           & 4.593                     & 2.960                 & 100\%                     & R                         & 59 nm                               &                                                        &                       &                           &                          & \multicolumn{1}{l}{}                \\
\multicolumn{11}{l}{Pyrolyzed carboxylate precursor E2 (0.5M)}                                                                                                                                                                                                                                                                                              \\
350           & \multicolumn{5}{c}{- amorphous -}                                                                                                               & \multicolumn{1}{l}{}                                   &                       & \multicolumn{1}{l}{}      &                          & \multicolumn{1}{l}{}                \\
400           & \multicolumn{5}{c}{- amorphous -}                                                                                                               & \multicolumn{1}{c}{}                                   &                       & \multicolumn{1}{c}{}      & \multicolumn{1}{c}{}     &                                     \\
450           & 3.789                     & 9.489                 & 100\%                     & A                         & 8 nm                                &                                                        &                       &                           & \multicolumn{1}{c}{}     &                                     \\
550           & 3.788                     & 9.517                 & 77\%                      & A                         & 20 nm                               & 4.597                                                  & 2.960                 & 23\%                      & \multicolumn{1}{c}{R}    & 27 nm                               \\
650           & 4.595                     & 2.961                 & 81\%                      & R                         & 46 nm                               & 3.785                                                  & 9.520                 & 19\%                      & \multicolumn{1}{c}{A}    & 33 nm                               \\
750           & 4.595                     & 2.962                 & 100\%                     & R                         & 57 nm                               &                                                        & \multicolumn{1}{r}{}  &                           &                          & \multicolumn{1}{l}{}                \\
\hline  
              & \multicolumn{1}{l}{}      & \multicolumn{1}{l}{}  & \multicolumn{1}{l}{}      & \multicolumn{1}{l}{}      & \multicolumn{1}{l}{}                & \multicolumn{1}{l}{}                                   & \multicolumn{1}{l}{}  & \multicolumn{1}{l}{}      &                          & \multicolumn{1}{l}{}             
\end{tabular*}
\label{tbl:xrd_all}
\end{table*}

Gravimetric analysis confirmed 0.15M titanium loading in precursor E1 and 0.5M in E2, correponding to the efficiency of cation exchange 45\% and 100\% respectively. Higher ratio of aqueous to organic phase volumes (5:1 vs 3:1) achieves complete extraction and efficient loading of metal cations in the organic phase.

As evident from thermal analysis the formation mechanism of nanoparticles from produced organometallic precursors differ substantially from the standard hydrolytic sol-gel process. The details of the phase content, stability and surface functional groups of produced nanoparticles can be derived from the XRD and FTIR analysis.

\subsection{XRD analysis}

The precipitate (P) from standard "one pot" sol-gel system was used as precursor to produce TiO$_2$ nanoparticles by calcination. The XRD patterns of calcined nanopowders
are shown in Fig.~\ref{FIG:3}A and used as reference. The XRD pattern of the as dried precipitate does not show clear signs of crystallinity, indicating that the TiO$_2$ precursor before calcination is in amorphous state. The annealing temperature of 450 $^\circ$C induces the crystallization of precipitate and distinctive peaks emerge, attributable to the tetragonal structure (space group: I41/amd) of crystalline anatase. All the peaks are fully indexed within ICDD PDF Pattern 00-021-1272 indicating that anatase titania has been formed at this temperature. The results of the Rietvield refinement concerning phase content and crystallite size are reported in Table~\ref{tbl:xrd_all}.

Fig.~\ref{FIG:3}B shows the XRD patterns of powders produced by pyrolysis of the organometallic precursor (E1). The sample pyrolyzed at 400 $^\circ$C shows only a weak and diffuse peak at 2$\theta$ = 25.3$^\circ$ attributable to the [101] reflection of anatase, which demonstrates that the sample still remains in mostly amorphous state. Pyrolyzation at 450 $^\circ$C induces the formation of pure nano-crystalline anatase from this precursor. The corresponding diffractogram in Fig.~\ref{FIG:3}B closely reflects the XRD pattern of precipitate (P) calcined at 450 $^\circ$C (Fig.~\ref{FIG:3}A) indicating identical phase state. The crystallite size of the produced anatase nanoparticles is ca. 7-8 nm (Table~\ref{tbl:xrd_all}). 

When the processing temperature of the precursors is increased to 650 $^\circ$C weak [110] (27.4$^\circ$, 100\% intensity) and [101] (36.1$^\circ$) reflections appear in the XRD pattern of calcined precipitates corresponding to the tetragonal structure (space group: P42/mnm) of crystalline titania rutile polymorph, which marks the beginning of a high-temperature anatase-to-rutile transformation accompanied by rapid grain growth (Table~\ref{tbl:xrd_all}). The diffractograms of powders obtained by pyrolyzation of organometallic precursor at this temperature show additional reflections of rutile. All additional peaks are readily indexed within standard ICCD PDF Pattern 00-021-1276 for rutile. 

\begin{figure*}[h]
	\centering
		\includegraphics[width=0.9\linewidth]{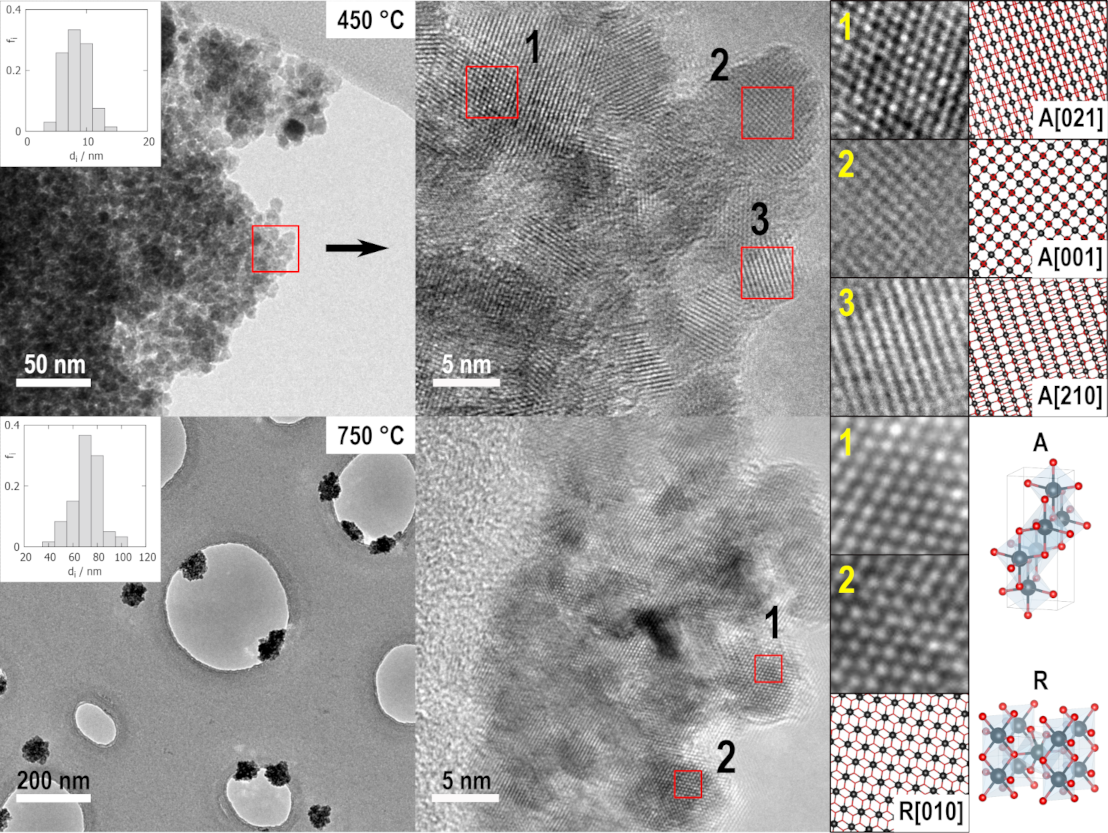}
	\caption{ TEM images of samples produced by pyrolysis of organometallic precursor at 450 $^\circ$C (E1-450, top row) and 750 $^\circ$C (E1-750, bottom row). Middle - magnified high-resolution TEM image of few ca. 10 nm anatase nanocrystals (top) and a secondary aggregate of crystallographically aligned and fused ca. 10 nm rutile grains (bottom). Right - lattice fringes of nanoparticles and corresponding projections of anatase (A) and rutile (R) crystal structures. }
	\label{FIG:4}
\end{figure*}

At 750 $^\circ$ both precursors complete the transformation from anatase to rutile and rutile becomes the predominant crystalline phase. This course of transformation is consistent with previous reports \cite{Zhang2002, Nicula2003}. No traces of brookite phase are found in any of the samples as confirmed by the absence of indicative [121] (30.8$^\circ$, 90\% intensity) reflection (ICDD PDF 00-029-1276).

At 750 $^\circ$C simultaneously with the anatase to rutile transformation the calcination of the precipitate obtained by the reference method finally leads to the crystallization of the impurity phase in P-750 (Fig.~\ref{FIG:3}A) identified as sodium titanate Na$_2$Ti$_6$O$_13$ (ICDD PDF 00-014-0277, space group: C12/m1). It is evident that an appreciable amount (ca. 20\%, Table~\ref{tbl:xrd_all}) of by-products of the standard monophasic "one-pot" synthesis are remaining in the produced materials despite extensive purification measures and the confirmed absence of chloride in washwater effluent. In this case sodium from the precipitating basic agent NaOH binds the chloride from the titanium chloride solution forming a large amount of sodium chloride in the precipitate, which is difficult to remove. The impurity phase becomes apparent in the XRD as the product of the high-temperature reaction between the unremoved residues of the sodium salt and titanium dioxide during calcination. In industrial pyro-hydrolysis process used to produce commercial TiO$_2$ the powder is treated with steam at 450-550 $^\circ$C to remove chlorine-containing groups. Our results show that within the non-hydrolytic extraction-pyrolysis method the purification is achieved already at the precursor preparation stage, employing a biphasic system, in which one of the phases accumulates the target product, Ti-containing organometallic compound, and the other accumulates by-products of the synthesis, in this case chloride anions.

Moreover, non-hydrolytic pyrolysis of organometallic extract allows facile phase control by just varying the temperature of the pyrolytic treatment. The samples E1-550 and E2-550 produced by pyrolysis at 550 $^\circ$C mimic the mixed phase content of Aeroxide P25, consisting of ca. 70-80 wt.\% anatase and 20-30 wt.\% rutile with ca. 20-27 nm particle size in both phases (Table~\ref{tbl:xrd_all}). The phase evolution of the particles is independent of the concentration of organometallic precursors, with just the start of the anatase-to-rutile conversion beginning at ca. 100 $^\circ$C lower temperature, which allows to use concentrated precursors to produce phase-pure anatase.

\subsection{TEM analysis}

Fig.~\ref{FIG:4} shows characteristic TEM snapshots of TiO$_2$ particles produced by pyrolyzation of organometallic precursor E1 at 450 $^\circ$C (E1-450). The particles are roughly spheroidal, which is characteristic for nucleation directly from the amorphous state. A tight size distribution is measured by particle counting with an average size of 8.3 nm in agreement with XRD assessment based on Rietvield modelling (Table~\ref{tbl:xrd_all}). High resolution TEM shows lattice fringes unambiguously identifiable as projections of anatase crystal structure.

As indicated by DSC analysis (Fig.~\ref{FIG:2}), at ca. 560 $^\circ$C the amorphous phase is fully expended and the primary crystallite formation terminates. Pyrolyzation at higher temperature is accompanied by primary crystallite fusion with simultaneous anatase-to-rutile transition. Fig.~\ref{FIG:4} shows the particles pyrolyzed at 750 $^\circ$C (E1-750). The highly irregular-shaped particles are constructed by joining of a number of ca. 10 nm primary crystalline grains forming secondary aggregates. According to strict thermodynamic arguments, single-phase anatase is only stable below ca. 11-14 nm size due to its slightly lower surface free energy \cite{Zhang1998}, with rutile being the macroscopically (ca. >35 nm) stable polymorph \cite{Zhang1998, Zhang2000, Zhu2005}. Epitaxial attachment of aggregated phase-pure anatase nanoparticles leads to joining on their facets, where the rutile nucleation is initiated assisted by intermediate phase, e.g. brookite \cite{Penn1998, Zhang1998b, Tossell1999} or high-pressure TiO$_2-II$ phase \cite{Zhu2015}, formation at twinned interface between contacting anatase grains \cite{Penn1998, Tossell1999, Zhang2000b, Oskam2003}. Rutile transformation then propagates rapidly into the bulk of the crystallites. 
The epitaxially fused nanoparticles with the rutile structure are noted in the high resolution TEM snapshots. Close inspection of TEM microphotographs (Fig.~\ref{FIG:4}) clearly reveals that the particles have many adjacent rutile crystallites sintered together and crystallographically aligned with respect to each other. The average size of the secondary aggregates is 70.3 nm, consistent with XRD analysis (Table~\ref{tbl:xrd_all}).

\subsection{Infrared (IR) spectroscopy}

While XRD provides the information about the bulk of the particles, FT-IR is a surface-sensitive method. It probes the the surface region and chemistry of the particles, which is not identified by XRD. 
\begin{figure}[h]
	\centering
		\includegraphics[width=0.8\linewidth]{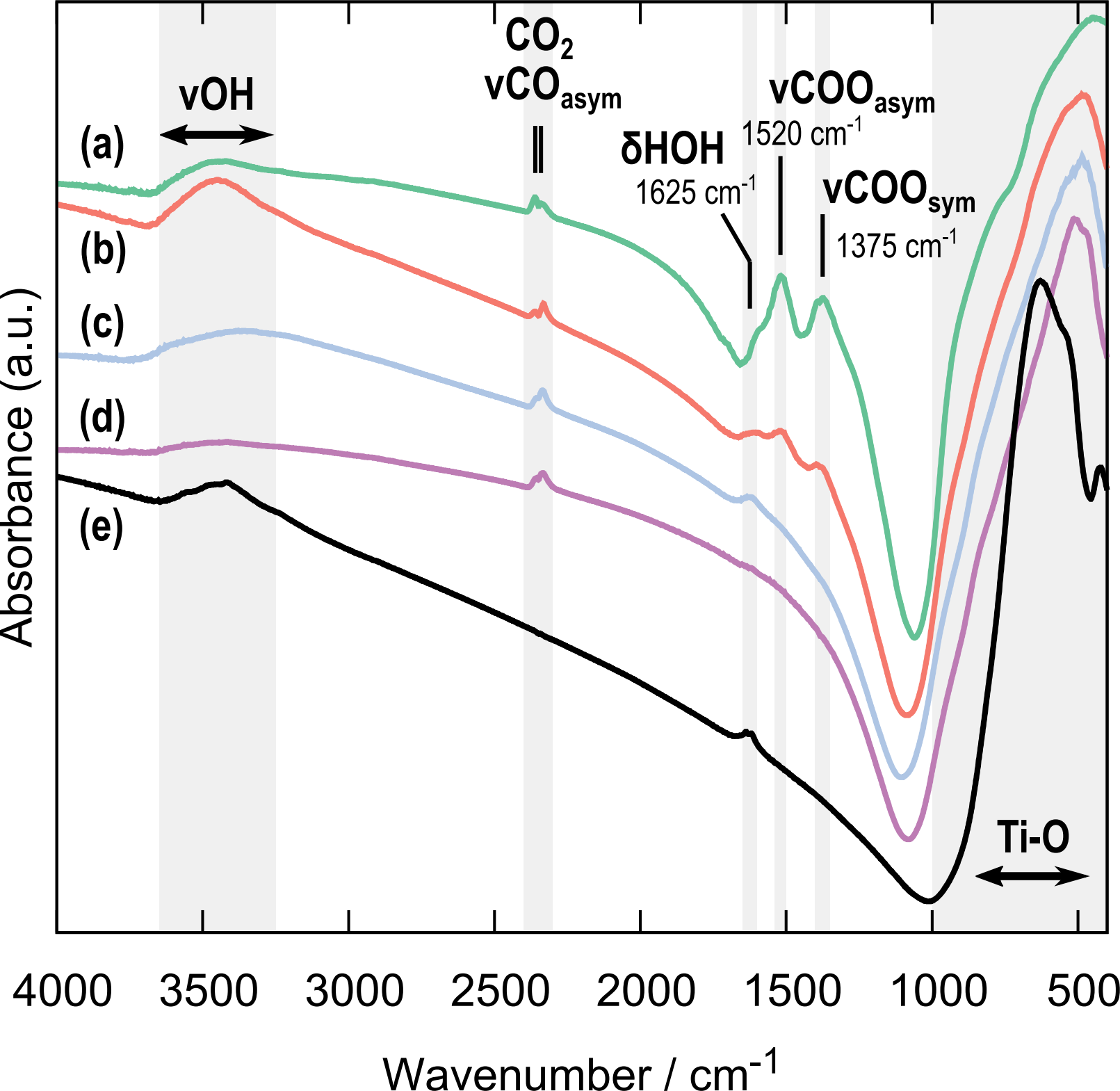}
	\caption{ IR spectra of TiO$_2$ samples produced by pyrolyzation of carboxylate-based Ti precursor: (a) - E1-350, (b) - E1-400, (c) E1-450, (d) E1-550 (complete elimination of surface residual organics is noted at 450 $^\circ$C); (e) - electro-dispersed sample calcined at 850 $^\circ$C (ED-850). }
	\label{FIG:5}
\end{figure}
FT-IR spectra of nanoparticles produced by pyrolyzation of carboxylate precursor (E1) are shown in Fig.~\ref{FIG:5}. The broad IR absorption band at 3200-3600 cm$^{-1}$ in all samples corresponds to the stretching vibration of the hydroxyl groups ($\nu$OH) terminated on TiO$_2$ surface (Ti-OH) and adsorbed molecular water. The spectrum band centered at 1625 cm$^{-1}$ is from the bending vibration of chemisorbed water $\delta$H-O-H. Weak absorbance doublet at 2341 cm$^{-1}$ and 2360 cm$^{-1}$ is due to asymmetric C-O stretching vibration of carbon dioxide. Since this band is present only in the pyrolyzed samples, we ascribe it to CO$_2$ residues from the oxidation of pentanoic acid captured on the pores in the nanopowder\footnote{cf. Fig.~\ref{FIG:5}e, showing FT-IR of nanoparticles produced by electrodispersion of initial metallic powder without any chemicals and calcined at high temperature (see below)}. Peaks at 1375 cm$^{-1}$ and 1520 cm$^{-1}$ represent
symmetric $\nu$COO$^-_{sym}$ and asymmetric $\nu$COO$^{-}_{asym}$ stretching modes of the carboxylate (COO$^-$) functional group \cite{Nolan2009} coming from the organic component of the precursors used in the synthesis process. The magnitude of the separation between carboxylate stretches $\Delta$ = $\nu$COO$^-_{asym}$ - $\nu$COO$^-_{sym}$=$\sim$145 cm$^{-1}$ is consistent with the value for ionic carboxylate complexes ($\sim$164 cm$^{-1}$ for ionic acetate \cite{Nakamoto2008}). This typically indicates that the carboxylate group is bound to the surface Ti-centers in bidentate bridging configuration \cite{Zelek2007}, where one metal cation is bound to one of the oxygens of the COO$^-$ group and another metal cation to the other oxygen. Carboxylate vibrations disappear with pyrolyzation temperature starting at 450 $^\circ$C, indicating complete elimination of organic residues when the anatase crystallization starts. A remaining broad and strong absorption band between 1000 and 400 cm$^{-1}$ envelops a set of peaks corresponding to the intrinsic Ti-O-Ti, O-Ti-O and Ti-O lattice vibrations of nano-crystalline titanium oxides \cite{cdc1982}.

\subsection{Defect-induced room-temperature ferromagnetism}

\begin{figure}[h]
	\centering
		\includegraphics[width=1\linewidth]{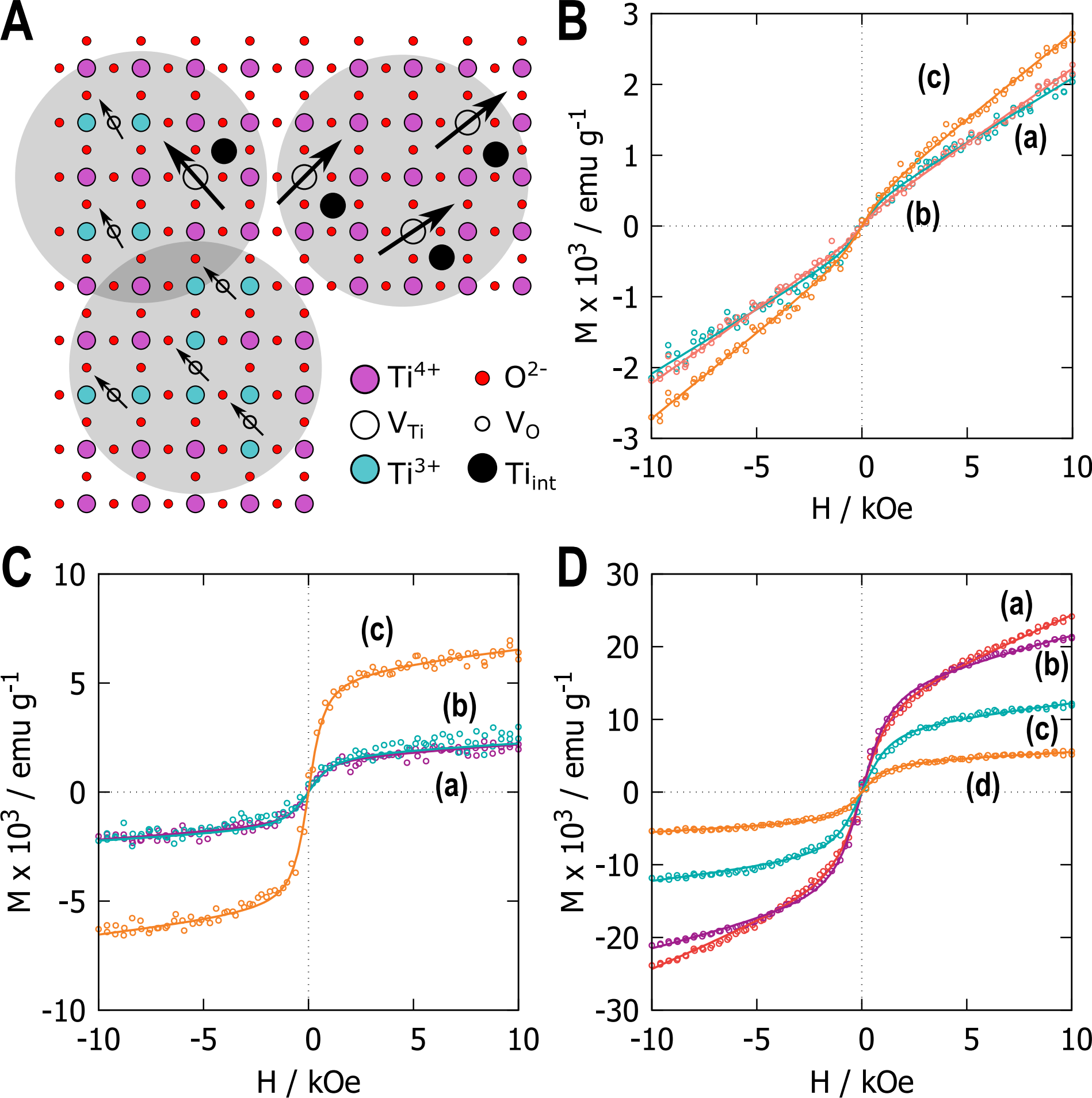}
	\caption{ Defect-induced room-temperature ferromagnetism in Ti-oxide nanoparticles: A - schematic polaron model of mesoscale ferromagnetic coupling between magnetic moments produced at cationic or anionic defect sites in stoichiometric TiO$_2$. M-H magnetization curves recorded from: B - precipitates in a monophasic system calcined at (a) 550 $^\circ$C (P-550, anatase phase), (b) 650 $^\circ$C (P-650), (c) 750 $^\circ$C (P-750, rutile); B - pyrolyzed Ti-carboxylate precursor at (a) 450 $^\circ$C (E1-450, anatase), (b) 550 $^\circ$C (E1-550, anatase), (c) 750 $^\circ$C (E1-750, rutile); C - electrically dispersed nanoparticles calcined at (a) 350 $^\circ$C (ED-350), (b) 450 $^\circ$C (ED-450), (c) 550 $^\circ$C (ED-550), (d) 750 $^\circ$C (ED-750). }
	\label{FIG:6}
\end{figure}

To obtain an estimation of the defect content in the produced materials we characterize their ferromagnetic behaviour by VSM. The recorded magnetization curves M-H are shown in Fig.~\ref{FIG:6} with the magnetic field varying in the range between -10 kOe to +10 kOe. TiO$_2$ precipitates produced by standard sol-gel are predominantly paramagnetic $M\approx\chi H$ (Fig.~\ref{FIG:6}B), irrespective of the calcination temperature. The mass magnetic susceptibility $\chi$ of the samples calcined at 550 $^\circ$C (P-550) and 650 $^\circ$C (P-650) having anatase as the dominant phase is ca. $2\times 10^{-7}$ cm$^3$ g$^{-1}$. The anatase-to-rutile transition in the calcined samples is accompanied just by a slight increase in $\chi$ at 750 $^\circ$C (P-750, Table~\ref{tbl:magnetic}).

\begin{table}[h]
\small
\caption{Summary of magnetic properties. The polaron dipole moment $\mu$ in units of Bohr magneton $\mu_B$ = 9.27 $\dots \times 10^{-21}$ erg/G.}
\begin{tabular*}{0.48\textwidth}{@{\extracolsep{\fill}}lrrrrr}
\hline
\multicolumn{1}{c}{T, $^\circ$C} & \multicolumn{1}{c}{$\chi$ / $10^{-7}$ }     & \multicolumn{1}{c}{$M_{sat}$ / $10^{-3}$}   & \multicolumn{1}{c}{$N$ / $10^{13}$} & \multicolumn{1}{c}{$\mu\mu_B^{-1}$ / $10^3$}  \\
\multicolumn{1}{c}{}             & \multicolumn{1}{c}{cm$^3$ g$^{-1}$}  & \multicolumn{1}{c}{emu g$^{-1}$}  & \multicolumn{1}{c}{g$^{-1}$}      & \multicolumn{1}{c}{}       \\
\hline
\multicolumn{5}{l}{Sol-gel reference sample (P)} \\
550                              & 1.8                    & 0.3                  & 0.3                      & 12                        \\
650                              & 2.1                    & 0.1                  & 0.1                      & 15                        \\
750                              & 2.4                    & 0.3                  & 0.4                      & 9.7                          \\
\multicolumn{5}{l}{Pyrolyzed carboxylate precursor (E1)} \\
450                              & 0.6                    & 1.6                  & 1.5                      & 12                          \\
550                              & 0.5                    & 1.8                  & 2.0                      & 9.7                          \\
750                              & 1.3                    & 4.8                  & 4.0                      & 13                          \\
\multicolumn{5}{l}{Electrodispersed sample (ED)} \\
350                              & 12.0                    & 13                  & 16                       & 8.6                          \\
450                              & 6.8                    & 15.5                 & 19                       & 8.6                          \\
550                              & 3.2                    & 9.5                  & 12                       & 8.6                          \\
750                              & 1.4                    & 4.3                  & 4.8                      & 9.7                          \\
                                 \hline                              
\end{tabular*}
\label{tbl:magnetic}
\end{table}

In contrast, the M-H curves of powders produced by pyrolyzation of organometallic precursor (Fig.~\ref{FIG:6}C) show distinct room-temperature ferromagnetism superposed by the paramagnetic process. For all recorded magnetization curves the hysteresis is negligible with small values of coercitive field ranged within 50-90 Oe and the remanence factor varying around 5-8\%, indicating soft superparamagnetic-like behaviour characteristic for d0 ferromagnetism \cite{Esquinazi2020}.

Stoichiometric TiO$_2$, having all Ti in 4+ oxidation state is non-ferromagnetic due to the lack of unpaired electrons. Likewise, no magnetic impurities are introduced during the preparation or handling of the samples. The staring metallic Ti powder is chemically pure titanium as confirmed by X-Ray fluorescence. The content of magnetic impurities, i.e. Fe, Ni or Co, is below 0.1 at.\%. The doping of the produced
particles by the trace elements potentially present in the precursors is discarded as the origin of room-temperature ferromagnetism, because all samples were produced from the same metallic powder. Moreover, the absence of impurities in the sample produced by pyrolyzation of organometallic precursor is assured, on account of high selectivity of pentanoic acid extractant towards titanium ions and unfavorable
conditions for the concurrent bi-phasic extraction of other metallics in precursor preparation. Hence, room temperature ferromagnetism (RTFM) in the pyrolyzed samples is due to the intrinsic defects in the nanoparticles.

The origin of RTFM, known also as d0 magnetism, in undoped semiconducting metal oxides (SMO) is ascribed to intrinsic defects in the crystal lattice, generated during the production process \cite{Coey2005}. The emergence of ferromagnetism is typically accompanied by a high number of oxygen vacancies V$_O$ in the anionic sublattice of TiO$_2$ films \cite{Yoon2006, Yoon2007, Rumaiz2007, Liu2016}, nanoparticles \cite{Zhao2008, Wang2014} and single crystals \cite{Liu2019, Liu2020}. Likewise, the defect complex Ti$^{3+}$-V$_O$ was noted, which forms when oxygen is removed: the charge imbalance makes unpaired excess electrons occupy 3d state of nearby Ti ions, generating reduced Ti$^{3+}$ ions and providing the local magnetic moments \cite{Zhou2009, Santara2013}. In oxygen-defficient TiO$_{2-\delta}$ nanoparticles the reduction mechanism Ti$^{4+}$ $\rightarrow$ Ti$^{3+}$ can also be initiated by Ti interstitials, without involving oxygen vacancies \cite{Parras2013}. Moreover, cationic vacancies V$_{Ti}$ were shown to produce d0-ferromagnetism \cite{Peng2009, Wang2015}. The balance of these defects is strongly dependent on the preparation method.

\begin{figure}[t]
	\centering
		\includegraphics[width=1\linewidth]{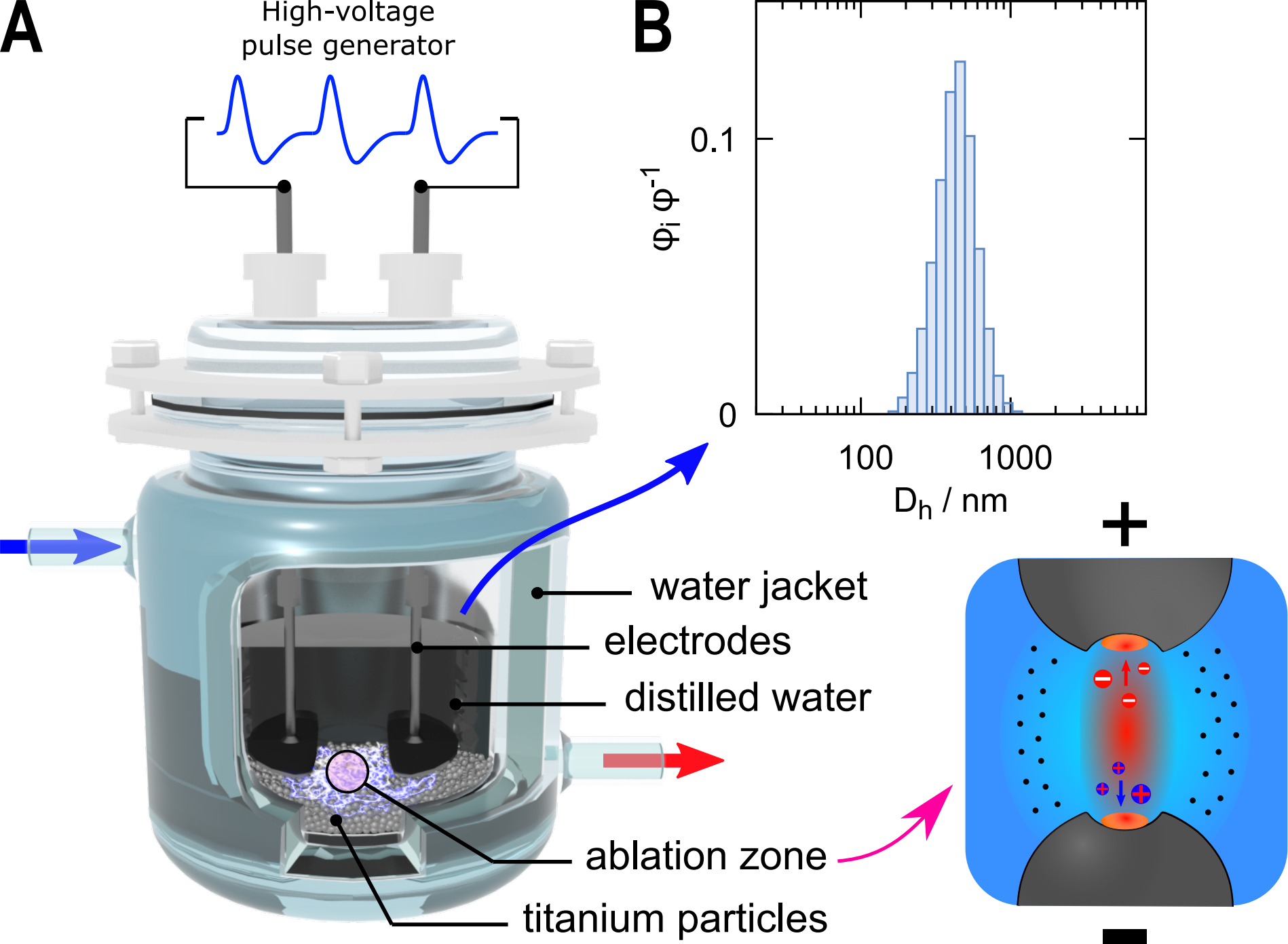}
	\caption{ Electric ablation experiment to produce highly defected Ti-oxide nanoparticles: A - schematic of the electric dispersion apparatus with the discharge ablation zone between a pair of electrodes pressed into the bed of titanium particles in distilled water. The repeated discharges ablate material from the surface of the particle bed; B - particle size distribution in as prepared colloid measured by DLS. }
	\label{FIG:7}
\end{figure}

In view of this, we have produced a reference sample with high defect loading by physical condensation - direct electric ablation of the starting metallic titanium precursor without additional chemicals, using a device shown schematically in Fig.~\ref{FIG:7}A \cite{Maiorov2019}. 
\begin{figure}[h]
	\centering
		\includegraphics[width=1\linewidth]{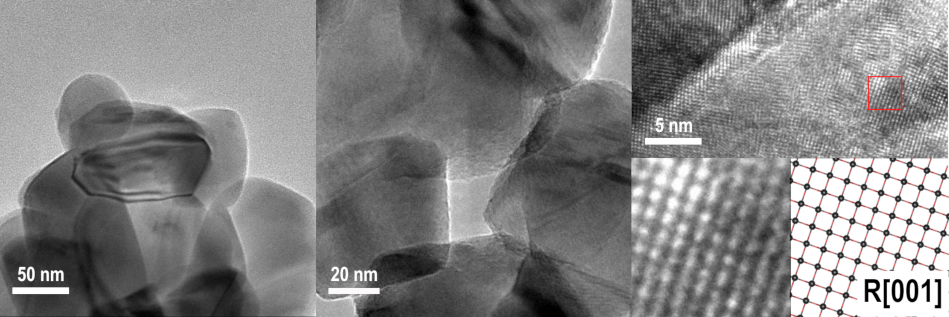}
	\caption{ TEM snapshots of electrically dispersed nanoparticles calcined at 850 $^\circ$C with varying magnification. HR-TEM image of lattice fringes and corresponding rutile lattice projection. }
	\label{FIG:8}
\end{figure}
\begin{figure}[h]
	\centering
		\includegraphics[width=1\linewidth]{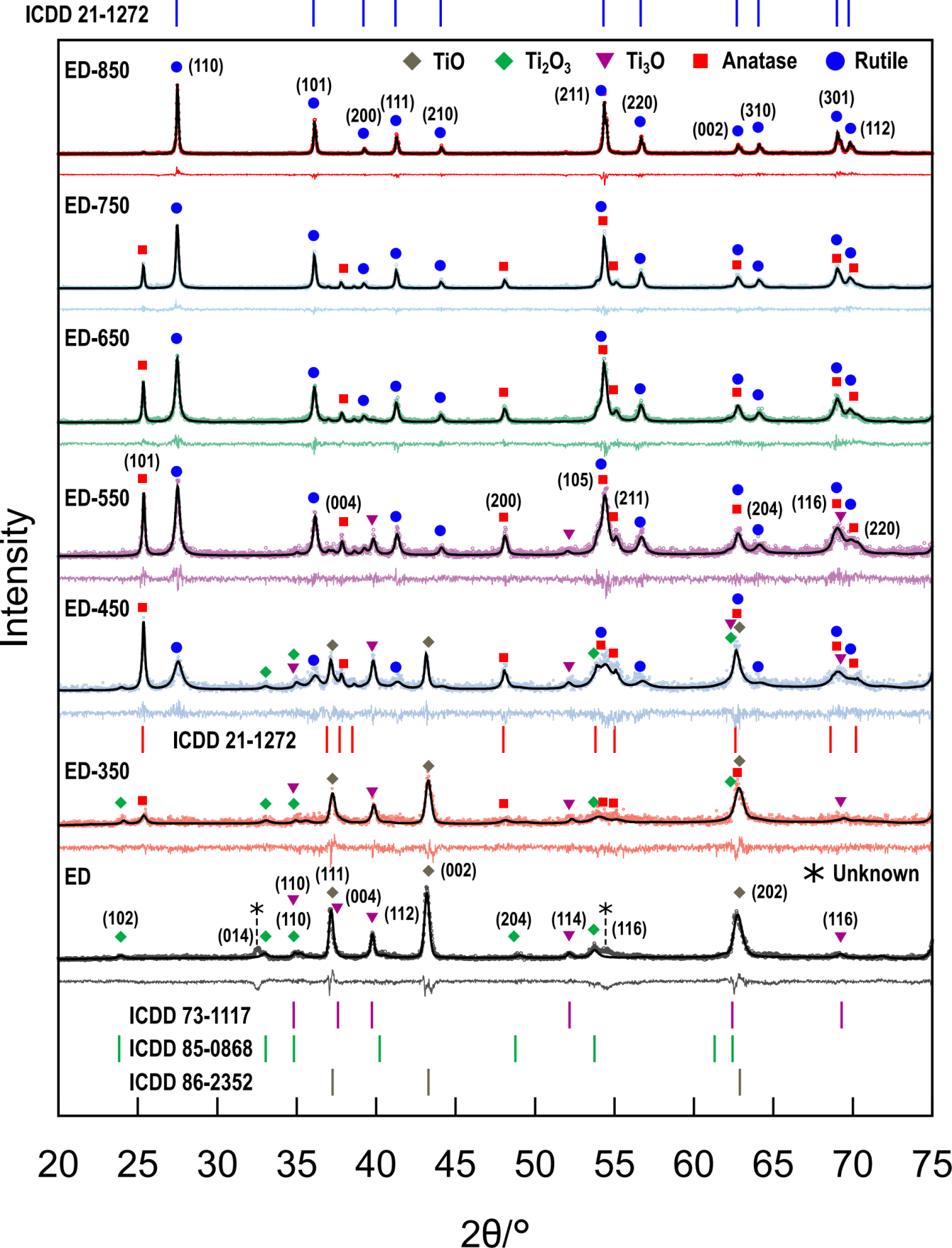}
	\caption{ XRD patterns of electrodispersed nanoparticles (ED) and phase content evolution under calcination treatment.}
	\label{FIG:9}
\end{figure}
\begin{table}[h]
\small
\caption{Summary of electrodispersed sample phase structure refinement. A - anatase, R - rutile.}
\begin{tabular*}{0.48\textwidth}{@{\extracolsep{\fill}}lrrrlr}
\hline
\multicolumn{1}{c}{T, $^\circ$C} & \multicolumn{2}{c}{Unit cell, \AA} & \multicolumn{1}{c}{wt.\%} & \multicolumn{1}{c}{Phase} & \multicolumn{1}{c}{Crystallite}     \\ \cline{2-3}
\multicolumn{1}{c}{}             & \multicolumn{1}{c}{a,b}  & \multicolumn{1}{c}{c}  & \multicolumn{1}{c}{}      & \multicolumn{1}{c}{}      & \multicolumn{1}{c}{size, d$_{XRD}$} \\
\hline
r.t.                             & 4.190                    & 4.190                  & 62\%                      & TiO                       & 26 nm                               \\
                                 & 5.092                    & 13.811                 & 25\%                      & Ti$_2$O$_3$                     & 12 nm                               \\
                                 & 5.148                    & 9.580                  & 13\%                      & Ti$_3$O                      & 35 nm                               \\
350                              & 4.187                    & 4.187                  & 53\%                      & TiO                       & 23 nm                               \\
                                 & 5.066                    & 13.803                 & 19\%                      & Ti$_2$O$_3$                     & 12 nm                               \\
                                 & 5.147                    & 9.548                  & 16\%                      & Ti$_3$O                      & 26 nm                               \\
                                 & 3.787                    & 9.428                  & 12\%                      & A                         & 24 nm                               \\
450                              & 4.589                    & 2.961                  & 45\%                      & R                         & 13 nm                               \\
                                 & 3.785                    & 9.526                  & 23\%                      & A                         & 52 nm                               \\
                                 & 4.195                    & 4.195                  & 14\%                      & TiO                       & 37 nm                               \\
                                 & 5.134                    & 9.626                  & 9\%                       & Ti$_3$O                      & 28 nm                               \\
                                 & 5.149                    & 13.642                 & 9\%                       & Ti$_2$O$_3$                     & 12 nm                               \\
550                              & 4.593                    & 2.958                  & 71\%                      & R                         & 27 nm                               \\
                                 & 3.784                    & 9.519                  & 22\%                      & A                         & 49 nm                               \\
                                 & 5.124                    & 9.659                  & 7\%                       & Ti$_3$O                      & 27 nm                               \\
650                              & 4.596                    & 2.961                  & 81\%                      & R                         & 33 nm                               \\
                                 & 3.786                    & 9.526                  & 18\%                      & A                         & 52 nm                               \\
                                 & 5.137                    & 9.695                  & below 5\%                 & Ti$_3$O                      & 10 nm                               \\
750                              & 4.596                    & 2.961                  & 86\%                      & R                         & 52 nm                               \\
                                 & 3.786                    & 9.527                  & 14\%                      & A                         & 62 nm                               \\
850                              & 4.593                    & 2.960                  & $\sim$100\%               & R                         & 73 nm                               \\
                                 & 3.786                    & 9.491                  & below 5\%                 & A                         & 28 nm \\
                                 \hline                              
\end{tabular*}
\label{tbl:xrd_spark}
\end{table}
Briefly, a 20 g portion of metallic titanium powder was added to 200 ml of distilled water forming a particle bed. Flat titanium electrodes connected to a spectrometric discharge generator operating at 100 Hz and 2 kV were pressed onto the particle bed. After turning on the generator a luminous network of intermittent sparks appears within the particle bed. Reactive dispersion of titanium using distilled water as the working dielectric is accompanied by rapid ejection and oxidation of vaporized titanium and the formation of titanium oxide nanoparticles. The rapid \textit{in situ} quenching leads to a highly defected oxygen-deficient structure.

The initially transparent liquid turned completely black after 30 min of device operation, evidencing the condensation of nanodispersed phase. Dynamic light scattering (DLS, Malvern Instruments, Zetasizer Nano S90) (Fig.~\ref{FIG:7}B) showed that the hydrodynamic particle size in the decanted supernatant varies in a broad range between ca. 100 nm to 1 $\mu$m, which indicates that the particles form associates in liquid
medium due to the lack of stabilization \cite{Segal2008, Zablotskaya2009, Segal2010, Zablotskaya2018, Zablotskaya2015}. The produced suspension was dried, the electrodispersed nanopowder (ED) was calcined at 350 $^\circ$C (ED-350), 450 $^\circ$C (ED-450), 550 $^\circ$C (ED-550), 650 $^\circ$C (ED-650), 750 $^\circ$C (ED-750), 850 $^\circ$C (ED-850) in static air for 2 h to sample phase structure transformation and ferromagnetic properties.

The phase content of electrodispersed nanoparticles is non-stoichiometric oxygen-deficient oxides (Fig.~\ref{FIG:9}, Table~\ref{tbl:xrd_spark}), predominantly titanium monoxide TiO (space group: Fm-3m, indexed within ICDD PDF 01-086-2352), which is able to accommodate up to 10-20\% of O vacancies in the cubic structure \cite{Banus1972}, but also trigonal Ti$_2$O$_3$ structure (space group: R3c, ICDD PDF 01-085-0868) and cubic Ti$_3$O structure (space group: P-31c, ICDD PDF 01-073-1117) have been found with average crystallite size ca. 25-35 nm. Anatase appears after calcination at 350 $^\circ$C and the content of non-stoichiometric oxides begins to decrease due to a more complete oxidation. A mixed phase sample ED-750 of stoichiometric anatase (14 wt.\%) and rutile TiO$_2$ (86 wt.\%) is finally obtained calcining at 750 $^\circ$C with grain size ca. 50-60 nm. Higher calcination temperature (850 $^\circ$C) completes the anatase-to-rutile transition with particles having slightly irregular rounded shape (Fig.~\ref{FIG:8}) and the lattice fringes are fully indexed to projections of rutile structure.

As expected, due to non-stoichiometry and a high defect content generated by the very rapid reactive quenching of electrically ablated vapourized material in surrounding water, the produced nanoparticles show significantly enhanced ferromagnetic properties (Fig.~\ref{FIG:6}D).

Theoretical studies indicate that concentrated cationic \cite{Peng2009} and anionic \cite{Lu2011} defects couple ferromagnetically by exchange interaction, forming bound magnetic polarons (BMPs), which are bubbles of ferromagnetially ordered spins (Fig.~\ref{FIG:6}A) \cite{Coey2005}. More quantitative understanding of the polaron-induced magnetism can be obtained within the BMP-model by fitting the measured magnetization with the equation of the type \cite{Coey2005, Santara2013, Zablotsky2020, Zablotsky2020b}
\begin{equation}
M=M_{sat}\mathcal{L}\left(\xi\right)+\chi H \label{eq:Msat}
\end{equation}
where the first term describes the superparamagnetic contribution and the second term describes to first order of magnetic field H the paramagnetic processes in non-ferromagnetic matrix, which are not related to the superparamagnetic properties of the sample. Here, $M_{sat}$ is the saturation magnetization [emu g$^{-1}$], which is the product of the number of bound magnetic polarons involved in the polarization process per unit mass of the sample $N$ [g$^{-1}$] and $\mu$ - the spontaneous dipole moment [erg G$^{-1}$] of a polaron; $\mathcal{L}\left(x\right)=\coth\left(x\right)-\frac{1}{x}$ is the Langevin function, $\xi = \frac{\mu H}{k_B T}$ - Langevin parameter, $k_B = 1.38 \ldots \times 10^{-16}$ erg K$^{-1}$ is the Boltzmann constant, T - temperature (ca. 300 K) and $\chi$ is paramagnetic susceptibility [cm$^3$ g$^{-1}$]. We have analyzed the recorded M-H curves for all samples in terms of the bound polaron model Eq.~\eqref{eq:Msat} using polaron concentration $N$, their dipole moment $\mu$ and magnetic susceptibility $\chi$ as fitting parameters. The theoretical lines closely follow the measured M-H curves for all samples (Fig.~\ref{FIG:6}). The calculated parameters are summarized in Table~\ref{tbl:magnetic}.

The superparamagnetic saturation magnetization of samples produced by pyrolyzation of organometallic precursor is about an order of magnitude higher than for reference sol-gel particles, indicating a more defected structure. $M_{sat}$ consistently increases with the pyrolyzation temperature, reaching 4.8 memu/g at 750 $^\circ$C (E1-750).

The highest superparamagnetic saturation $M_{sat}$ =15.5 memu/g is obtained for electrodispersed sample ED-450, indicating highly defected structure was produced. Calcination at high temperature >450 $^\circ$C leads to the decrease of ferromagnetic properties due to the gradually diminishing defect content, such as the filling of oxygen vacancies in static air. The remaining superparamagnetic saturation at the highest annealing temperature 750 $^\circ$C is $M_{sat}$ = 4.3 memu/g.

The phase composition, crystallite size (ca. 50-60 nm) and polaron content ($\sim$ 4 - 4.8 $\times 10^{13}$ g$^{-1}$) in both rutile-rich samples E1-750 and ED-750 (Table~\ref{tbl:magnetic}) produced respectively by pyrolyzation and electrodispersion is
similar. It is worth noting that according to the TGA traces (Fig.~\ref{FIG:2}) showing continuous mass gain, the pyrolyzation of
organometallic precursor relies on oxygen absorption from the environment, which might favor formation of oxygen vacancies. This is in contrast to the precipitated sample, which shows similar phase composition and crystallite size, but an order of magnitude smaller polaron concentration, indicating a defect-free crystal lattice, since the stoichiometry of the precipitates in aqueous sol-gel is assured on account of the hydrolysis and condensation process.

Previously, 10 nm TiO$_{2-\delta}$ nanoparticles with mostly anatase phase produced by sol-gel showed saturation magnetization
9.5 memu/g and coercitivity of 96 Oe after annealing under reducing H$_2$/Ar atmosphere. The origin of the ferromagnetism was attributed to a large content of V$_O$. This was contested by Parras et al. \cite{Parras2013}, where Magneli phase TiO$_{1.84}$ was stabilized under similar conditions. Rather than the influence of oxygen vacancies, the saturation magnetization 10 memu/g was instead ascribed to the Ti$^{3+}$ ions induced by
reduction, initiated by formation of Ti interstitials. Polycrystalline TiO$_2$ treated under 4 MeV Ar$^{5+}$ ion irradiation showed 0.4 memu/g saturation with 168 Oe coercitive field\cite{Sanyal2013}. Oxygen vacancies V$_O$ were the dominant defects after irradiation. Wang et al. \cite{Wang2014b} produced 10-20 nm anatase nanoparticles by sol-gel and reported $\sim$1 memu/g saturation and 93 Oe coercitive field attributed to V$_O$ content and charge transfer interactions between Ti$^{3+}$ and Ti$^{2+}$. Recently \cite{Akshay2018, Chanda2018}, sol-gel produced 10 nm single-phase anatase TiO$_2$ calcined at
400 $^\circ$C showed ca. 0.5-13 memu/g saturation and 11-305 Oe coercitivity, on account of V$_O$. Wang et al. \cite{Wang2015} produced highly defected 10-20 nm anatase TiO$_2$ nanoparticles with 25 memu/g superparamagnetic saturation and 18 Oe coercitivity using solvothermal reaction. Moreover, defected TiO$_2$ showed remarkably higher photocatalytic activity vs. normal TiO$_2$ because of the more efficient charge transfer \cite{Wang2015}.

Our results show that the effective polaron magnetic moment $\mu\sim 10^4\mu_B$ ($\sim 10\times10^{-17}$ emu, Table~\ref{tbl:magnetic}) does not distinctly depend on the preparation method or thermal treatment. Remarkably, very similar magnitudes of polaron magnetic moments were obtained in two previous investigations, in which this quantity was estimated: $\mu\sim 6.5-8.9\times 10^{-17}$ emu was measured in large nanorods\cite{Santara2013} with diameter $\sim$100 nm and length of few $\mu$m produced by NaOH-assisted solvothermal processing with vacuum annealing and/or low-pressure calcination. The phase content varied from pure brookite to anatase-rutile mix with large concentration of V$_O$ and Ti$^{3+}$ states. The same $\mu =7 - 7.5 \times 10^3\mu_B$ polaron moment was measured in sol-gel produced 10 nm single-phase anatase\cite{Akshay2018} due to V$_O$, despite saturation magnetization being roughly 2 orders of magnitude smaller than in the first study \cite{Santara2013}. Our results show that the polaron properties are roughly unaffected neither by the preparation method, phase content of the produced TiO$_2$, nor annealing conditions, which lead just to the variability of polaron concentration $N$ in the samples.

A unambiguous one-to-one linear correlation exists between the concentration of bound magnetic polarons $N$ and the oxygen vacancy content\cite{Santara2013}. Theoretical calculations indicate that a single oxygen vacancy V$_O$ can produce magnetic moment of about 1.4-2.4 $\mu_B$ \cite{Sanyal2013, Liu2019, Liu2020}, cation vacancy V$_{Ti}$ – about 2.3-4 $\mu_B$ \cite{Peng2009, Sanyal2013, Liu2019, Liu2020} and di-vacancy – 2 $\mu_B$ \cite{Peng2009, Esquinazi2020}. For a characteristic moment $\sim$ 2 $\mu_B$ per defect, ca. 0.4-0.7\% of defective cells are found in the electrodispersed powders with defect content decreasing with higher calcination temperature. In contrast, in pyrolyzed samples the defect content is ca. 0.01-0.04\% without an obvious dependence on the pyrolyzation temperature.

\section{Conclusions}

An organometallic precursor based on titanium carboxylate complex has been developed to produce phase-pure anatase, rutile and mixed phase TiO$_2$ nanoparticles with controllable polymorph content using a simple non-hydrolytic thermal process. The new approach, based on a different reaction mechanism, eliminates from consideration many reaction parameters, which are difficult to control in the hydrolytic sol-gel processes, particularly, such as instantaneous Ti:H$_2$O ratio, hydrolysis rate, pH, reaction modulators, and allows better control of the reaction process. A broad range of samples with varying structures, sizes and phase compositions has been produced by pyrolyzation of the carboxylate precursor. The temperature of the pyrolytic treatment comes out as a single parameter determining the grain size and crystal phase tuning of the final product in a simple, predictable and reproducible way.

The carboxylate precursor produced by biphasic cation exchange is long-term storable, hydrolytically stable and not sensitive to moisture contained in the air, unlike inorganic Ti chlorides or organometallic alkoxides commonly employed in the standard methods, which undergo vigorous hydrolysis. On this account titanium carboxylates are also ideal precursors for film deposition. The high chemical purity of the produced materials is assured on account of the biphasic extraction involved in the preparation of the precursor. Common impurities characteristic for the standard "one-pot" production methods have not been found during the investigation. The results are correlated against the traditional sol-gel method and confirm that the reported process is a good technique to produce TiO$_2$ nanoparticles having surfaces free from organics.

An advantage of the the carboxylate precursor in pyrogenic synthesis is that it allows to avoid the commonly used titanium chlorides and the related corrosive hazards in chlorine rich flames. Flame synthesis \cite{Ma2010, Manuputty2019} is a widely employed to produce high quality metal oxide nanoparticles through a one-step process. Hence, the carboxylate precursor is a favourable alternative in the production of TiO$_2$ nanoparticles.

\section*{Acknowledgements}
We acknowledge technical staff from the University of Latvia, which was involved in the preparation of some of the samples under grant agreement No. 1.1.1.1/16/A/085 (2017-2018).

%%%END OF MAIN TEXT%%%

%The \balance command can be used to balance the columns on the final page if desired. It should be placed anywhere within the first column of the last page.

\balance

%If notes are included in your references you can change the title from 'References' to 'Notes and references' using the following command:
%\renewcommand\refname{Notes and references}

%%%REFERENCES%%%
\bibliography{rsc} %You need to replace "rsc" on this line with the name of your .bib file
\bibliographystyle{rsc} %the RSC's .bst file

\end{document}